\documentstyle[12pt]{article}
\catcode`\@=11
\long\def\@makefntext#1{
\protect\noindent \hbox to 3.2pt {\hskip-.9pt
$^{{\ninerm\@thefnmark}}$\hfil}#1\hfill}		

 \def\@makefnmark{\hbox to 0pt{$^{\@thefnmark}$\hss}}  

\def\ps@myheadings{\let\@mkboth\@gobbletwo
\def\@oddhead{\hbox{}
\rightmark\hfil\ninerm\thepage}
\def\@oddfoot{}\def\@evenhead{\ninerm\thepage\hfil
\leftmark\hbox{}}\def\@evenfoot{}
\def\sectionmark##1{}\def\subsectionmark##1{}}

\newcounter{sectionc}\newcounter{subsectionc}\newcounter{subsubsectionc}
\renewcommand{\section}[1] {\vspace{0.6cm}\addtocounter{sectionc}{1}
\setcounter{subsectionc}{0}\setcounter{subsubsectionc}{0}\noindent
	{\bf\thesectionc. #1}\par\vspace{0.4cm}}
\renewcommand{\subsection}[1] {\vspace{0.6cm}\addtocounter{subsectionc}{1}
	\setcounter{subsubsectionc}{0}\noindent
	{\it\thesectionc.\thesubsectionc. #1}\par\vspace{0.4cm}}
\renewcommand{\subsubsection}[1] {\vspace{0.6cm}\addtocounter{subsubsectionc}{1}
	\noindent {\rm\thesectionc.\thesubsectionc.\thesubsubsectionc.
	#1}\par\vspace{0.4cm}}

\newcounter{appendixc}
\newcounter{subappendixc}[appendixc]
\newcounter{subsubappendixc}[subappendixc]

\renewcommand{\appendix}[1] {\vspace{0.6cm}
        \refstepcounter{appendixc}
        \setcounter{figure}{0}
        \setcounter{table}{0}
        \setcounter{equation}{0}
        \renewcommand{\thefigure}{\Alph{appendixc}.\arabic{figure}}
        \renewcommand{\thetable}{\Alph{appendixc}.\arabic{table}}
        \renewcommand{\theappendixc}{\Alph{appendixc}}
        \renewcommand{\theequation}{\Alph{appendixc}.\arabic{equation}}
        \noindent{\bf Appendix \theappendixc #1}\par\vspace{0.4cm}}

\renewenvironment{thebibliography}[1]
	{\begin{list}{\arabic{enumi}.}
	{\usecounter{enumi}\setlength{\parsep}{0pt}
\setlength{\leftmargin 1.25cm}{\rightmargin 0pt}
	 \setlength{\itemsep}{0pt} \settowidth
	{\labelwidth}{#1.}\sloppy}}{\end{list}}

\newcounter{itemlistc}
\newcounter{romanlistc}
\newcounter{alphlistc}
\newcounter{arabiclistc}

\newcommand{\fcaption}[1]{
        \refstepcounter{figure}
        \setbox\@tempboxa = \hbox{\tenrm Fig.~\thefigure. #1}
        \ifdim \wd\@tempboxa > 6in
           {\begin{center}
        \parbox{6in}{\tenrm\baselineskip=12pt Fig.~\thefigure. #1}
            \end{center}}
        \else
             {\begin{center}
             {\tenrm Fig.~\thefigure. #1}
              \end{center}}
        \fi}

\newcommand{\tcaption}[1]{
        \refstepcounter{table}
        \setbox\@tempboxa = \hbox{\tenrm Table~\thetable. #1}
        \ifdim \wd\@tempboxa > 6in
           {\begin{center}
        \parbox{6in}{\tenrm\baselineskip=12pt Table~\thetable. #1}
            \end{center}}
        \else
             {\begin{center}
             {\tenrm Table~\thetable. #1}
              \end{center}}
        \fi}

\def\@citex[#1]#2{\if@filesw\immediate\write\@auxout
	{\string\citation{#2}}\fi
\def\@citea{}\@cite{\@for\@citeb:=#2\do
	{\@citea\def\@citea{,}\@ifundefined
	{b@\@citeb}{{\bf ?}\@warning
	{Citation `\@citeb' on page \thepage \space undefined}}
	{\csname b@\@citeb\endcsname}}}{#1}}

\newif\if@cghi
\def\cite{\@cghitrue\@ifnextchar [{\@tempswatrue
	\@citex}{\@tempswafalse\@citex[]}}
\def\citelow{\@cghifalse\@ifnextchar [{\@tempswatrue
	\@citex}{\@tempswafalse\@citex[]}}
\def\@cite#1#2{{$\null^{#1}$\if@tempswa\typeout
	{IJCGA warning: optional citation argument
	ignored: `#2'} \fi}}

\def\fnt#1#2{\footnotetext{\kern-.3em
	{$^{\mbox{\sevenrm #1}}$}{#2}}}

 1
 1
 1

\font\tenrm=cmr10

\font\ninerm=cmr9

\newcommand{\be}{\begin{equation}}
\newcommand{\ee}{\end{equation}}
\newcommand{\bea}{\begin{eqnarray}}
\newcommand{\eea}{\end{eqnarray}}
\def\KB{\overline{K}^0}
\def\PT{\widetilde P}
\begin{document}

\hfill{
\begin{tabular}{r}
Universit\'a di Napoli Preprint DSF-34/96 (June 1996)\\
hep-ph/9608236
\end{tabular}
}

\vspace{.5cm}

\begin{center}
{\bf NON-LEPTONIC TWO-BODY WEAK DECAYS OF CHARMED MESONS
AND CP-VIOLATING ASYMMETRIES\\}

\vspace*{1cm}
PIETRO SANTORELLI \\
{\it Dipartimento di Scienze Fisiche, Universit\`a ``FEDERICO II''
di Napoli,\\
and \\
INFN, Sezione di Napoli\\
Mostra D'Oltremare Pad. 19-20, 80125 Napoli, Italy\\}
\end{center}

\vspace*{1cm}
\begin{abstracts}
{\small The non-leptonic two body decays of D mesons are studied in the 
framework of an improved factorization approximation. The final state 
interaction effects are taken into account assuming them dominated by 
nearby resonances. The agreement with experimental data is quite good. 
CP-violating asymmetries are predicted.}
\end{abstracts}

\section{Introduction}

A theoretical description of exclusive non-leptonic decays of charmed 
mesons based on first principles has not yet been achieved. Although the 
short-distance effects due to hard gluon exchange can be resummed and 
the effective hamiltonian has been constructed at next-to-leading 
order,\cite{MarBur} the evaluation of its matrix elements requires
non-perturbative techniques. In this respect, a classical analysis 
based on QCD sum rules has been presented in three papers by 
Blok and Shifman.\cite{BlokShif} However only the general trends were 
reproduced by their analysis, while no agreement with current 
experimental data was obtained. More recently, 
Martinelli and collaborators\cite{Marti} 
have proposed a procedure to study two-body non-leptonic weak 
decays in numerical simulation of lattice QCD. 
Since no numerical result have been obtained as yet, one has however 
to resort to models.

We here present one such model\cite{BLMPS} based on the factorization
approximation with annihilation terms and rescattering effects due to
the resonances coupled to the final states, that has been rather
successful to account for the experimental data about two-body decays of
charged and neutral D mesons in PP and PV final states. This feature
has made it possible for us to obtain reliable predictions for
the related CP-violating asymmetries.


\section{Weak Decay Amplitudes}

The effective weak hamiltonian for Cabibbo allowed 
non-leptonic decays of charmed particles is given by 
\bea
H_{eff}^{\Delta C = \Delta S} = 
{G_F \over \sqrt 2}\,V_{ud}\,V_{cs}^*\;
&&\!\!\!\!\!\!\!\!\!
\Bigl[C_2\;\bar{s}^\alpha\,\gamma_\mu\,(1-\gamma_5)\,c_\alpha\;
\bar{u}^\beta\,\gamma^\mu\,(1-\gamma_5)\,d_\beta + \nonumber\\
&&\!\!\!\!\!\!\!\!\!
C_1\;\bar{u}^\alpha\,\gamma_\mu\,(1-\gamma_5)\,c_\alpha\;
       \bar{s}^\beta\,\gamma^\mu\,(1-\gamma_5)\,d_\beta \Bigr] + 
      {\rm h.c.}\,\,\,.
\eea
For $\Delta C = -\Delta S$ processes the hamiltonian is obtained from 
the previous equation with the substitution $s \leftrightarrow d$.
The effective weak hamiltonian for Cabibbo-first-forbidden (CFF) 
non-leptonic decays reads
\bea
H_{eff}^{\Delta C =\pm 1, \Delta S=0} &=& 
       {G_F \over \sqrt 2}\Bigl\{\,V_{ud}\,V_{cd}^*\;
      \Bigl[C_1\;Q_1^d\;+\;C_2\;Q_2^d \Bigr] + \,V_{us}\,V_{cs}^*\;
      \Bigl[C_1\;Q_1^s\;+\;C_2\;Q_2^s \Bigr] \nonumber\\ 
 &  &  -\;   \,V_{ub}\,V_{cb}^*
      \; \sum_{i=3}^6 \;C_i\; Q_i \Bigr\}\;\;\;+\;\; {\rm h.c.} \,\,\,.
\label{e:HeffCf}
\eea
In Eq. (\ref{e:HeffCf}) the operators are defined as 
\bea
Q_1^d & = & 
       \bar{u}^\alpha\,\gamma_\mu\,(1-\gamma_5)\,d_\beta\;
       \bar{d}^\beta\,\gamma^\mu\,(1-\gamma_5)\,c_\alpha, \nonumber\\
Q_2^d & = & 
       \bar{u}^\alpha\,\gamma_\mu\,(1-\gamma_5)\,d_\alpha\;
       \bar{d}^\beta\,\gamma^\mu\,(1-\gamma_5)\,c_\beta, \nonumber\\
Q_3   & = & 
       \bar{u}^\alpha\,\gamma_\mu\,(1-\gamma_5)\,c_\alpha\;
      \sum_q\; \bar{q}^\beta\,\gamma^\mu\,(1-\gamma_5)\,q_\beta, \nonumber\\
Q_4   & = & 
       \bar{u}^\alpha\,\gamma_\mu\,(1-\gamma_5)\,c_\beta \;
       \sum_q\;\bar{q}^\beta\,\gamma^\mu\,(1-\gamma_5)\,q_\alpha, \nonumber\\
Q_5   & = & 
       \bar{u}^\alpha\,\gamma_\mu\,(1-\gamma_5)\,c_\alpha\;
       \sum_q\;\bar{q}^\beta\,\gamma^\mu\,(1+\gamma_5)\,q_\beta, \nonumber\\
Q_6   & = & 
       \bar{u}^\alpha\,\gamma_\mu\,(1-\gamma_5)\,c_\beta \;
       \sum_q\;\bar{q}^\beta\,\gamma^\mu\,(1+\gamma_5)\,q_\alpha\;.
\eea
The operator $Q_1^s$ ($Q_2^s$) in Eq. (\ref{e:HeffCf}) is obtained from 
$Q_1^d$ ($Q_2^d$) with the substitution $(d \to s)$. $\alpha$ and $\beta$ 
are colour indices (that we will omit in Eq. (\ref{e:diver})) 
and in the ``penguin'' operators $q$ ($\bar{q}$) is to be summed over 
all active flavors ($u$, $d$, $s$).

For the Wilson coefficients we used the anomalous dimension matrices 
calculated at next-to-leading order.\cite{MarBur} Assuming 
$\Lambda_4^{\overline{MS}} = 300~MeV$, at the scale 
$\mu = 1.5~GeV$ for the ``scheme independent prescription'' (cfr
Buras {\it et al.} in Ref.~\cite{MarBur}) we 
obtain $C_1 = -0.628$, $C_2 = 1.347$, $C_3 = 0.027$,
$C_4 = -0.057$, $C_5 = 0.015$, $C_6 = -0.070$. 

In the factorization approximation the matrix elements of $H_{eff}$
are written in terms of matrix elements of {\it currents},
$(V_{q^{\prime}}^q)^{\mu}$ = $\bar{q}^{\prime}\,\gamma^{\mu}\,q$
and $(A_{q^{\prime}}^q)^{\mu}$ = $\bar{q}^{\prime}\,\gamma^{\mu}\gamma_5\,q$.
To evaluate these matrix elements,\footnote{The matrix elements of axial
vector current and axial density between the vacuum and $\eta
(\eta^{\prime})$ state are evaluated following the authors of
Ref. \cite{Diakonov}.}~~we adopt the usual definition of the decay
constants and the form factors.\cite{BSW} 
The $q^2$ dependence of the involved form factors ($f_1(q^2),~f_0(q^2)$ and 
$A_0(q^2)$ for PP and PV final states) is assumed to be dominated
by the nearest resonances. The values $f_1(0)$ and $f_0(0)$ are fixed
by using SU(3) symmetry and the semileptonic decay rate $D^0 \to K^- e^+ \nu$. 
Since the data on $D$ meson decays show large SU(3) breaking effects,
in our fit we allowed $a_{cs} (\equiv A_0^{c\to s}(0))$ to be different by 
$a_{cd} = a_{cu} (\equiv A_0^{c\to d(u)}(0))$ 
and the values obtained by the fit are $a_{cs} = 0.59$,
and $a_{cd} = 1$.\footnote{A direct QCD sum rule calculation\cite{CDS} of
$A_0(q^2)$ shows the $q^2$ dependence compatible with the pole, but a 
different SU(3)-breaking effects: $a_{cs}/a_{cu}=1.10 \pm 0.05$ at 
$q^2=0$.}

In the W-exchange and annihilation terms, however, the large and time-like
$q^2$ values needed, together with the suggested existence of resonances
with masses near to the $D$-meson mass, make a prediction based on the
lightest mass singularity unjustified. These terms depend on the matrix 
elements of current divergences between the vacuum and two-meson states. 
We write them, with the help of the equations of motion, in the 
way indicated in the following examples:
\bea
\!\!\!\!\!\!  <K^-\pi^+|\partial^\mu(V_s^d)_\mu|0>\!\!\! &=&\!\!\! 
i\,(m_s-m_d)\,<K^-\pi^+|\,\bar{s}d\,|0> 
\equiv  i\,(m_s-m_d)\,{M_D^2 \over f_D}\,W_{PP}, \nonumber\\
\!\!\!\!\!\!\!\!\!\!\!\! \!\!\!\!\!\!\!\!  
\!\!\!\!\!\!\!\!\!\!\!\! \!\!\!\!\!\!\!\!  
\!\!\!\!\!\!\!\!\!\!\!\! \!\!\!\!\!\!\!\!  
\!\!\!\!\!\!\!\!\!\!\!\! \!\!\!\!\!\!\!\!  
< K^-\rho^+|\partial^\mu(A_s^d)_\mu|0\! > \!\!\!\!\!  &=&\!\!\! 
i(m_s+m_d)<\!\! K^- \rho^+|\,\bar{s}\gamma_5 d\,|0\! > \equiv 
-(m_s+m_d){2\,M_\rho \over f_D}\,\epsilon^*\!
\cdot\!\! p_K W_{PV}.
\label{e:diver}
\eea
We use SU(3) symmetry for the matrix elements of scalar and
pseudoscalar densities, and express all of them in terms of $W_{PP}$,
$W_{PV}$, which are, in our approach, free parameters of the fit.
Their magnitude turns out to be considerably larger than what one would
obtain assuming form factors dominated by the pole of the lightest
scalar or pseudoscalar meson, i.e. $K_0^*(1430)$ or $K(497)$. 


\section{Final State Interaction Effects}

As far as final state interactions (FSI) are concerned, we assume that
they are dominated by resonant contributions, and we neglect the
phase-shifts in exotic channels.\cite{FSI} In the mass region of
pseudoscalar charmed particles there is evidence\cite{PDG} for a
$J^P=0^-$ $\widetilde K(1830)$ (with $\Gamma = 250$ MeV and an observed
decay to $K \phi$) and a $J^P=0^-$ $\widetilde\pi(1770)$ with $\Gamma =
310$ MeV. These resonances have the right quantum numbers to construct
an $0^-$ octect which can couple to PV final states. The hamiltonian is
determined from charge conjugation and SU(3) symmetry. Analogously, the
FSI for $D\to PP$ decays should be dominated by the $J^P=0^+$ octect.
There is evidence for the existence of $\widetilde K^{*0}$ (with mass
1945$\pm$10$\pm$20 MeV, width 201$\pm$34$\pm$79 MeV and 52$\pm$14\%
branching ratio in $K \pi$). Unfortunately, no $a_0$ isovector resonance
has been observed up to now in the interesting mass region. However, we
assume his existence and fix the mass with an equispacing formula. 

The description of rescattering effects for Cabibbo forbidden $D^0$ decays 
is complicated by the presence of a coupling with a yet unobserved 
$f_0$ and $f_0^{\prime}$ isoscalar resonances, which should be 
singlet-octet mixtures.\footnote{Note that for the PV final state 
the C symmetry forbids the coupling of the final state to a singlet part 
of an hypotethical $\eta$-resonance. Thus, we need to fit
only a single phase, $\delta$.}~~In order to reduce the number of free
parameters we assume the scalar resonances behave as the tensor mesons
($J^P = 2^+$), $f_2(1270)$ and $f_2^{\prime}(1525)~$.\footnote{ The
$f_2^{\prime}$ is very weakly coupled to $\pi \pi$, and the $f_2$ has in
turn a small coupling to $K \overline{K}$.} This procedure relates the
coupling constant in the strong hamiltonian with the mixing angle $\phi$
between the singlet-octect part of $f_0$ and $f_0^{\prime}$.\footnote{For
further details see Ref.~\cite{BLMPS}.} 

In our model the FSI effect modifies the amplitudes, ${\cal A}_{\rm w}$, 
in the following way:~\footnote{An analogous expression holds for PP final 
state.}
\be
{\cal A} (D \to V_h\,P_k) = {\cal A}_{\rm w}
  (D \to V_h\,P_k) + c_{hk}[\exp(i\delta_8)-1]
   \sum_{h\prime k\prime} c_{h\prime k\prime}\,{\cal A}_{\rm w}
  (D \to V_{h\prime}\,P_{k\prime})
\label{e:res}
\ee
In Eq. (\ref{e:res}) $c_{hk}$ are the normalized ($\sum c_{hk}^2 = 1$) 
couplings $\widetilde P P V$ and
\be
\sin \delta_8 \, \exp (i \delta_8) =
{\Gamma ({\PT} ) \over 2\,(M_{\PT} - M_D) - i\,\Gamma ({\PT} )}
\ee
where $\PT$ is the resonance appropriate to the decay channel considered
($\widetilde \pi$ or $\widetilde K$).



\section{Comparison with Experimental Data on Branching Ratios and 
Predicted CP-Violating Asymmetries}

Using our model to evaluate the weak amplitudes ${\cal A}_{\rm w}$ and 
modifying them with FSI effects we are able to write the rates for all 
Cabibbo-allowed two-body decays and for Cabibbo-forbidden $D^+$ and 
$D^+_s$ as functions of eleven free parameters. Their values are fixed 
with a fit to the experimental data. The total $\chi^2_T=90.0$ for 45 data 
($\chi^2/dof = 2.6 $): 25 are data points for Cabibbo-allowed decays 
($\chi^2= 61.8$), 12 for CFF $D^+$ and $D^+_s$  
($\chi^2 = 18.8$), 4 for CFF $D^0\to PP$ ($\chi^2 = 1.7)$, and
4 for CFF $D^0\to PV$ ($\chi^2 = 7.7$). In all cases the agreement 
with the data is quite good, but the Table 1 shows that our 
model needs an improvement in describing the decays in PV final
state.\footnote{~This is the starting point for the authors of
Ref.~\cite{BLP}.} 

\begin{table}[ht]
\caption{Partial $\chi^2$ for each class of D decays.}
\begin{center}
\begin{tabular}{|c|c|c|}
\hline\hline 
& & \\ [-.35cm] 
Decays & \# data & $ \chi^2$ \\ [.25cm]
\hline 
& & \\ [-.35cm] 
$D^0 \to PP$ & 8  & 8.44 \\ [.25cm]
\hline
& & \\ [-.35cm] 
$D^+ \to PP$ & 5 & 9.56 \\[.25cm]
\hline
& & \\ [-.35cm] 
$D^+_s \to PP$ &  4 & 8.79 \\ [.25cm]
\hline
& & \\ [-.35cm] 
$D^0 \to PV$ & 12 & 18.35 \\ [.25cm]
\hline
& & \\ [-.35cm] 
$D^+ \to PV$ &  8 & 29.55 \\ [.25cm]
\hline
& & \\ [-.35cm] 
$D^+_s \to PV$ &  8 & 15.35 \\[.25cm]
\hline\hline
\end{tabular}
\end{center}
\end{table}

It is well known that CP-violating effects show up in a decay process 
only if the decay amplitude is the sum of two different
parts, whose phases are made of a weak (CKM) and a strong 
(final state interaction) contribution. If $A_1$ and $A_2$ denote the 
generic two weak amplitude contributing to the $D\to f$ amplitude, the 
CP-violating asymmetry in the decay rates will be:
\be
a_{CP}
= {2\;\Im(A_1 A_2^*) \, \sin(\delta_2-\delta_1) \over 
|A_1|^2+|A_2|^2+2\;\Re(A_1A_2^*)\,\cos(\delta_2-\delta_1)}
\ee
where $\delta_i$ are strong phases. Now in the Cabibbo-first-forbidden D 
decays the penguin operators in Eq. (\ref{e:HeffCf}) provide the 
different phases of the weak amplitudes $A_1$ and $A_2$. 

Having obtained a quite good description of the rates we may give
reliable prediction on CP-violating asymmetries for $D^+$ and $D^0$
Cabibbo-first-forbidden decays. In Table 2 we report CP-violating
asymmetries in $10^{-3}$ unit; the central values are obtained choosing
$\rho = 0.2$, $\eta = 0.3$ and $V_{cb} = 0.040$. The errors result from
the variation of $\rho$ and $\eta$ in the one-sigma region obtained in
Ref.~\cite{Marti2}. 

\begin{table}[ht]
\caption{CP-violating decay asymmetries for some $D^+$ and $D^0$ Cabibbo 
forbidden decays.}
\vspace*{.2cm}
\begin{center}
\begin{tabular}{|l|c||l|c|}
\hline\hline
decay channel & 
${\displaystyle 10^3 \times a_{CP}}$ &
decay channel &
${\displaystyle 10^3 \times a_{CP}}$ \\ \hline
$D^+ \to \rho^0 \pi^+$ & $ -1.17 \pm 0.68 $ &
$D^+ \to \KB K^+ $ & $ -0.51 \pm 0.30$ \\ \hline
$D^+ \to \rho^+ \pi^0$ & $ +1.28 \pm 0.74$ & 
$D^0 \to \pi^0 \eta$ & $- 1.43 \pm 0.83$ \\ \hline
$D^0 \to K^{*0} \KB$ & $-0.67 \pm 0.39$ &
$D^0 \to \pi^0 \eta^{\prime}$ & $+ 0.98 \pm 0.57$ \\ \hline
$D^0 \to \overline{K}^{*0} K^0 $ & $-0.67 \pm 0.39$ &
$D^0 \to \eta \eta$ & $- 0.50 \pm 0.29$  \\ \hline
$D^0 \to K^{*+} K^-$ & $+0.038 \pm 0.022$ &
$D^0 \to \eta \eta^{\prime}$ & $-0.28 \pm 0.16$ \\ \hline 
$D^0 \to K^{*-} K^+$ & $+ 0.16 \pm 0.09$ &
$D^0 \to \pi^0 \pi^0$ & $+0.54 \pm 0.31$  \\ \hline
$D^0 \to \rho^+ \pi^-$ & $+0.37 \pm 0.22$ & 
$D^0 \to \pi^+ \pi^-$ & $- 0.02 \pm 0.01$ \\ \hline
$D^0 \to \rho^- \pi^+$ & $-0.36 \pm 0.21$ &      
$D^0 \to K^+ K^-$ & $- 0.13 \pm 0.08$ \\ \hline
& & $D^0 \to K^0 \KB$ & $+0.28 \pm 0.16$ \\ 
\hline\hline
\end{tabular}
\end{center}
\end{table}

As we can see in Table 2 large asymmetries ($\approx 10^{-3}$)
are predicted in our model; at this end large final state phase shifts and 
penguin contributions played the fundamental role.


\section{Acknowledgments}

I wish to thank F. Buccella, M. Lusignoli, G. Miele and A. Pugliese
for many useful discussions.


\end{document}